\documentclass[prb,twocolumn,floatfix,showpacs]{revtex4}
\usepackage{graphicx}
\begin{document}
\title{Unexpected non-Wigner behavior in level-spacing distributions of
 next-nearest-neighbor coupled \textit{XXZ} spin chains}

\author{Kazue Kudo}
\affiliation{Graduate School of Humanities and Sciences, Ochanomizu
University, 2-1-1 Ohtsuka, Bunkyo-ku, Tokyo 112-8610, Japan}
\email{kudo@degway.phys.ocha.ac.jp}
\author{Tetsuo Deguchi}
\affiliation{Department of Physics, Ochanomizu University, 2-1-1
Ohtsuka, Bunkyo-ku, Tokyo 112-8610, Japan}
\email{deguchi@phys.ocha.ac.jp}

\date{08 May 2003}

\begin{abstract}
 The level-spacing distributions of \textit{XXZ} spin chains 
 with next-nearest-neighbor couplings  are studied under periodic
 boundary conditions. We confirm that integrable \textit{XXZ} spin chains 
 mostly have the Poisson distribution as expected. On the contrary, 
  the level-spacing distributions of next-nearest-neighbor coupled 
\textit{XXZ} chains 
are given by non-Wigner distributions. It is against the expectations, 
since the models are nonintegrable.  
\end{abstract}


\maketitle

\section{\label{sec:intro}Introduction}

Random matrix theories have been successfully applied 
to the energy spectra of not only chaotic systems but also 
 quantum spin systems. 
\cite{Montam,Hsu,Poil,Pals,8vertex,Meyer,Meyer2,Angles,AM} 
If a given Hamiltonian is integrable by the Bethe ansatz, 
the level-spacing distribution should be  
described by the Poisson distribution: 
\begin{equation}
 P_{\rm Poi}(s) = \exp (-s).
 \label{eq:Poisson}
\end{equation} 
If it is  not integrable, the level-spacing distribution should be 
given by the Wigner distribution:
\begin{equation}
 P_{\rm Wig}(s) = \frac{\pi s}{2} \exp \left( - \frac{\pi s^2}{4}
					   \right).
 \label{eq:Wigner}
\end{equation}
These behaviors of the level-spacing distribution have been observed in
one-dimensional (1D),\cite{Hsu,Poil,Angles}
2D,\cite{Pals,Montam,Meyer2} and 3D\cite{Meyer2} spin systems.
They are also confirmed for strongly correlated systems,\cite{Alt}  
and applied to the recent study of quantum dots. \cite{Held}  

The numerical observations 
\cite{Hsu,Poil,Angles,Pals,Montam,Meyer2,Alt,Held} are important.
In fact, for the quantum spin systems, 
there has been no theoretical or analytical derivation of     
the suggested behaviors of the level-spacing distribution.  
Random matrix theories do not necessarily extend to them.  
However, it seems that no counterexample has been shown explicitly 
 for  quantum spin systems. 
 The suggested behaviors 
  have been numerically confirmed for quite a large number 
 of quantum spin systems and statistical lattice systems. 
 \cite{Hsu,Poil,Angles,Pals,Montam,Meyer2,AM}  
 Moreover, the behaviors  
are often considered as empirical rules, by which  we can 
practically determine whether a given lattice system  
is integrable or not \cite{AM}. Thus, it could be nontrivial 
if one would find such a quantum spin system that does not show  
the standard behaviors of the level-spacing distribution.  

Recently, it has been found that an extraordinary symmetry  appears 
for special cases of the integrable \textit{XXZ} spin chain: 
the \textit{XXZ} Hamiltonian commutes with the $sl_2$ loop algebra 
at some particular values of the \textit{XXZ} coupling associated with  
roots of unity. \cite{DFM}  
The loop algebra is an infinite-dimensional Lie algebra, 
which holds only at the particular values of 
 the \textit{XXZ} coupling  and  not for generic values. 
 It is suggested  numerically \cite{FM} 
 that the standard Bethe ansatz does not hold  for the special cases.  
Furthermore, the dimensions of degenerate eigenspaces  
of the loop algebra are given by some exponential functions 
of the system size, \cite{Deguchi}  
and they can be extremely large. It should therefore be nontrivial 
how the large degeneracies are resolved by 
nonintegrable perturbative terms in the spin Hamiltonian. 
 Thus, the $sl_2$ loop algebra symmetry may     
motivate us to reconsider not only the Bethe ansatz solvability 
of the integrable \textit{XXZ} chain 
but also the standard statistical behaviors of energy levels 
for various \textit{XXZ} chains close to the integrable one. We note that some
level crossings of the $sl_2$ loop algebra are shown in Fig.~1 of
Ref.~\onlinecite{KD}.

In this paper, we discuss level-spacing distributions for  
the spin $\frac12$ \textit{XXZ} chains on finite sites 
under periodic boundary conditions.  
We mainly discuss nonintegrable cases, although  
the present study  has  been originally motivated 
by the $sl_2$ loop algebra symmetry of the integrable \textit{XXZ} chain. 
We first confirm that integrable spin chains 
show the Poisson distribution 
for generic cases.  Here we exclude the special cases related to  
roots of unity. 
Then, we consider the next-nearest-neighbor (NNN) coupled Heisenberg 
spin chain (or \textit{XXX} spin chain), which is nonintegrable. 
 We now note that the \textit{XXX} chain with NNN couplings 
 has the Wigner distribution, as shown in Ref. \onlinecite{Poil}. 
 For the \textit{XXZ} spin chains with  NNN couplings, however, 
 we find that the level-spacing distributions are not 
given by  the standard Wigner distribution. 
 The observation should be nontrivial since the systems are nonintegrable.  
Finally, we discuss possible reasons 
why the non-Wigner distribution 
is obtained for the NNN coupled \textit{XXZ} spin chains.

\section{\label{sec:method}Numerical procedure}

The Hamiltonian matrices can be separated into some sectors; in each
sector, the eigenstates have the same quantum numbers. This is because the
eigenvalues with different symmetries are uncorrelated. The \textit{XXZ} chains
have three trivial symmetries, namely, reflections, translations,
and spin rotations around the $z$ axis. Therefore, to
desymmetrize the Hamiltonians, we consider
three quantum numbers: parities, momenta, and total $S^z$. We calculate
the eigenvalues of the largest sectors. The largest sector of each
Hamiltonian has 440 eigenvalues for the lattice size $L=16$. To
calculate the eigenvalues, we use standard numerical methods, which are
contained in the LAPACK library.

To find universal statistical properties of the Hamiltonians, one has to
deal with unfolded eigenvalues instead of raw eigenvalues. The
unfolded eigenvalues are renormalized values, whose local density of
states is
equal to unity everywhere in the spectrum. In this paper, the unfolded
eigenvalues $x_i$ are obtained from the raw eigenvalues $E_i$ in the
following method. Let us define the integrated density of states as
\begin{equation}
 n(E) = \sum _{i=1}^N \theta (E-E_i).
\end{equation}  
Here $\theta (E)$ is the step function and $N$ is the number of the
eigenvalues. We choose some points of coordinates: $(E_i, n(E_i))$ for
$i=1,21,41,\cdots,N$. The average of integrated density of states $\langle
n(E) \rangle$ is approximated by the spline interpolation through the
chosen points. The unfolded eigenvalues are defined as
\begin{equation}
 x_i = \langle n(E_i) \rangle .
\end{equation}
The level-spacing distributions are given by the probability function
$P(s)$, where $s=x_{i+1}-x_i$.

\section{\label{sec:NNN}Next-nearest-neighbor coupled \textit{XXZ} chain}

Let us introduce the spin $\frac12$ \textit{XXZ} chain with NNN couplings on $L$ sites
by 
\begin{eqnarray}
 \mathcal{H}&=&J_1 \sum_{l=1}^L 
  \left( S^x_l S^x_{l+1} + S^y_l S^y_{l+1} 
   + \Delta_1 S^z_l S^z_{l+1} \right) \\ \nonumber
 &+&J_2 \sum_{l=1}^L
 \left[ \alpha \left( S^x_l S^x_{l+2} + S^y_l S^y_{l+2} \right) 
 + \Delta_2 S^z_l S^z_{l+2} \right] ,
\label{eq:H}
\end{eqnarray} 
where $S^a= (1/2)\sigma^a$ and $(\sigma^x,\sigma^y,\sigma^z)$
are the Pauli matrices; periodic boundary conditions are imposed. For
simplicity, we put $J_1=1$ 
hereafter  in the paper. The Hamiltonian is
nonintegrable for $J_2 \neq 0$, 
while it is integrable for $J_2=0$.

\begin{figure}
\includegraphics[width=8cm]{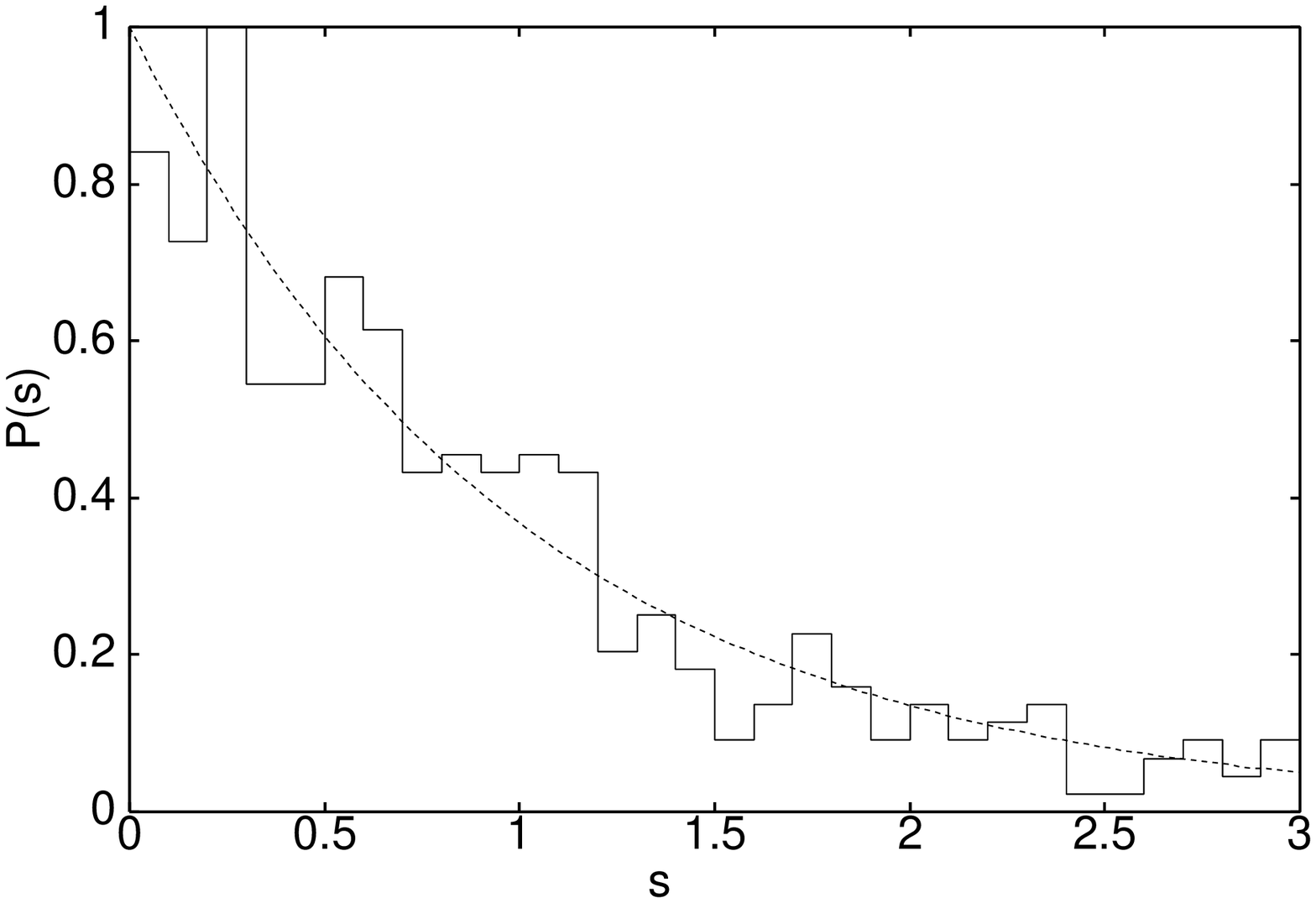}
\caption{\label{fig:in} Level-spacing distribution of the integrable \textit{XXZ}
 chain ($J_2=0$) for
 $L=16$, $J_1=1$, and $\Delta_1=0.1$ . 
 Broken line shows the
 Poisson distribution. There is no degeneracy at $s=0$. }
\end{figure}

Let us confirm the Poissonian behavior for the generic case of 
the integrable \textit{XXZ} spin chain. 
When $J_2=0$, the level-spacing distribution $P(s)$
mostly shows the Poisson distribution as shown in Fig.~\ref{fig:in}.
We confirmed numerically the standard result  
for some  generic values of the \textit{XXZ} coupling 
$\Delta_1$ ($0 \le \Delta_1 < 1$). We note that 
we exclude the special values of the \textit{XXZ} coupling that are given by 
 $\Delta_1=\cos(m \pi/N)$ for some integers $m$ and $N$, 
 where $\Delta_1$ is related to a root of unity, $q$, through 
the relation  $\Delta_1 = (q+1/q)/2$. \cite{DFM}

\begin{figure*}
\includegraphics[width=16cm]{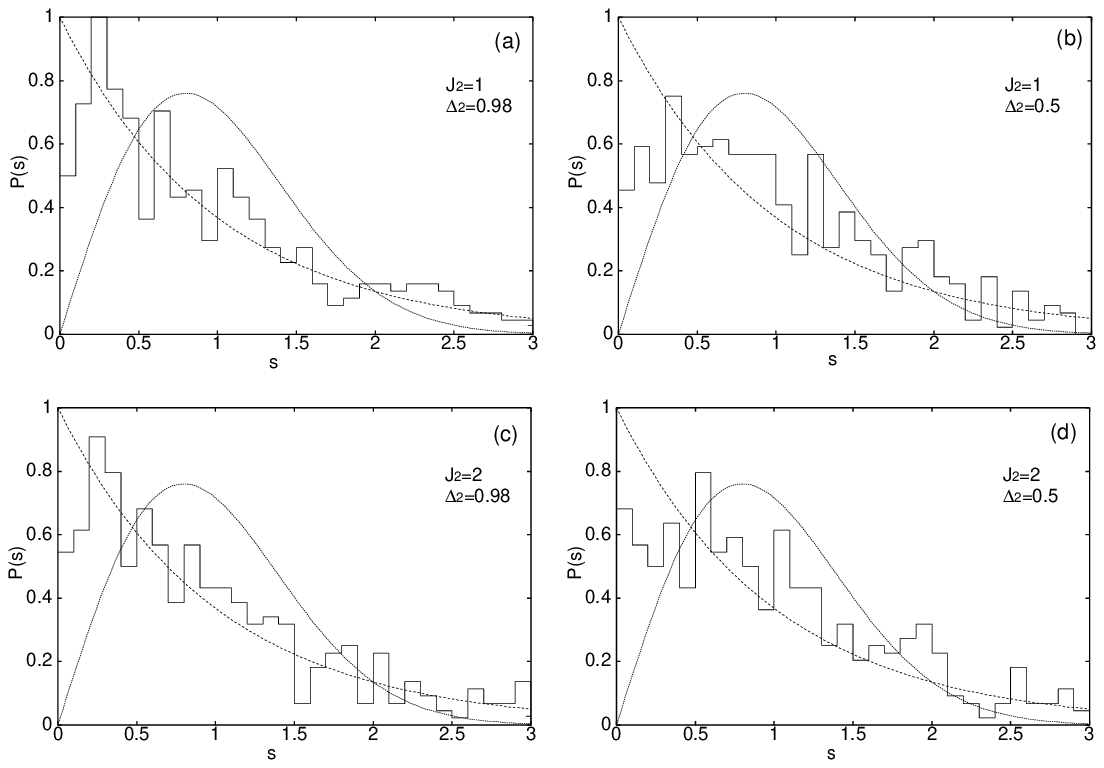}
\caption{\label{fig:nn} Level-spacing distributions of NNN coupled
 chain for $L=16$, $J_1 = \Delta_1 = \alpha =1$, (a) $J_2=1$, 
$\Delta_2 =0.98$;
 (b) $J_2=1$, $\Delta_2 =0.5$; (c) $J_2=2$, $\Delta_2 =0.98$; 
 (d) $J_2=2$, $\Delta_2 =0.5$. 
 Broken lines, the Poisson distribution; dotted lines, the Wigner
 distribution. There is no degeneracy: $P(s) =0$ at $s=0$  
  for all the four cases. }
\end{figure*}

In nonintegrable cases ($J_2\neq 0$), however, $P(s)$ shows a non-Wigner
distribution against expectations. 
The non-Wigner behavior of the NNN coupled \textit{XXZ} spin chains 
is generic.
We have calculated $P(s)$ for various 
combinations of $J_2$ and $\Delta_2$ when $\alpha=1$ and for various
values of $\Delta_2$ when $\alpha=0$. Here,  $\Delta_1=1$ 
for simplicity. When $\alpha=1$ and $\Delta_2=0.98$ 
, $P(s)$ 
looks similar  to  
 the Poisson distribution rather than the Wigner distribution for any
 value of $J_2$,  as
shown in Figs.~\ref{fig:nn}(a) and \ref{fig:nn}(c). 
 The Hamiltonian
is close to but not exactly the same as the Heisenberg chain, when
$\Delta_2=0.98$.  
We find such a Poisson-like distribution when  
$\Delta_2$ is very close to $1.0$ such as more 
than about $0.95$. For the NNN coupled Heisenberg chain ($\Delta_2=1$),
however, 
it was shown that $P(s)$ has the Wigner distribution. \cite{Hsu,Poil}

When $\Delta_2$ decreases, $P(s)$ becomes a non-Poisson
distribution as shown in Figs.~\ref{fig:nn}(b) and \ref{fig:nn}(d). 
However, it is not a Wigner-like distribution, 
either. The level-spacing distribution $P(s)$ strongly depends 
 on $\Delta_2$. On the other hand, it seems that
$P(s)$ is almost  independent of $J_2$. 

When only one parameter is 
nonzero, namely, for the case of either $\alpha=0$ or $\Delta_2=0$, 
the nonzero parameter does not 
change  $P(s)$ very much. In that case, 
$P(s)$ is given by neither a Poisson-like nor Wigner-like
distribution. 

\begin{figure}
\includegraphics[width=8cm]{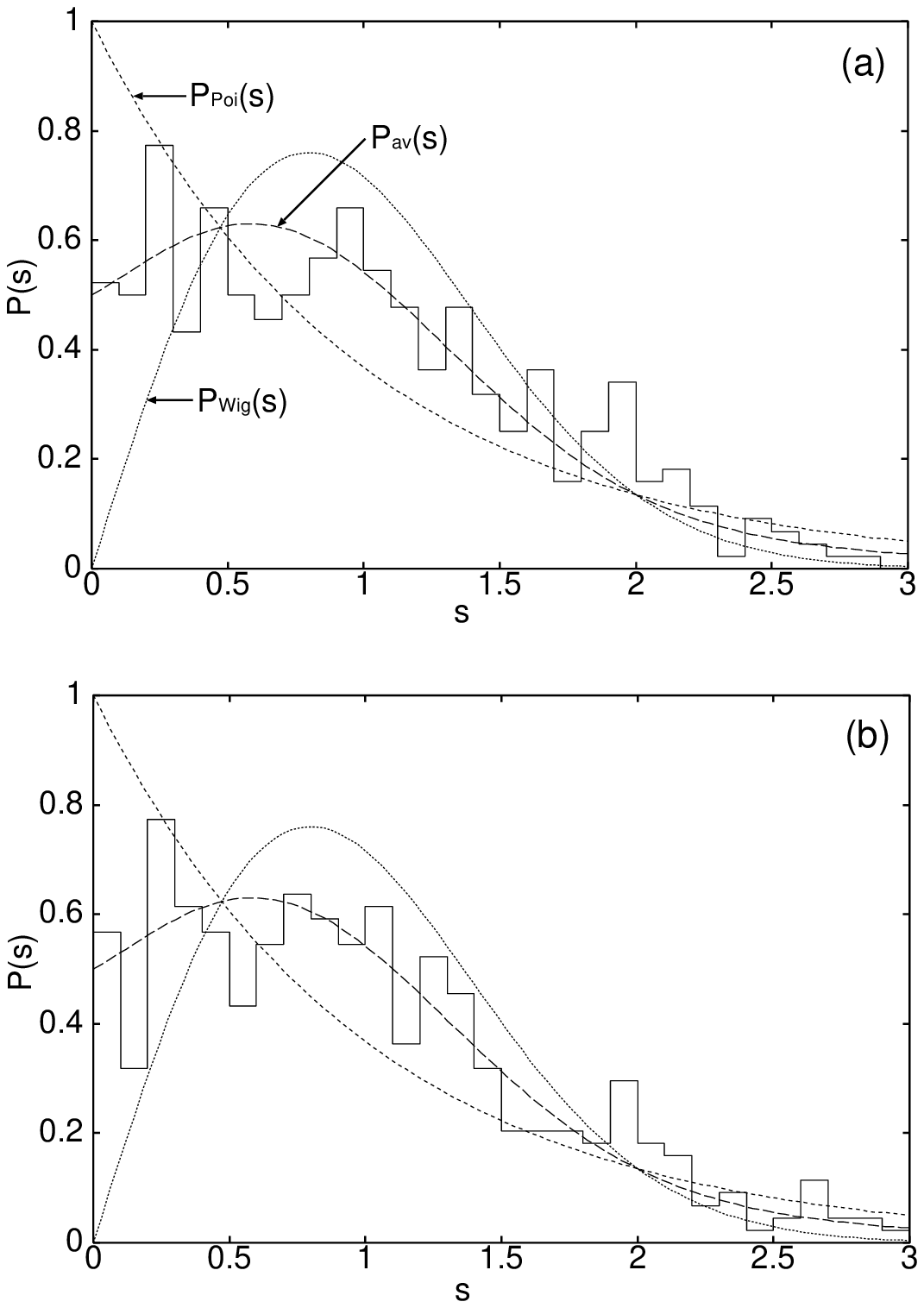}
\caption{\label{fig:nn2} Level-spacing distributions of NNN coupled 
chain for
 $L=16$, $J_1=\Delta_1=J_2=1$, (a) $\alpha=0$, $\Delta_2 =1$; (b) $\alpha=1$, 
$\Delta_2 =0$.
 Broken lines, the Poisson distribution; dotted lines, the Wigner
 distribution; long dashed lines, the arithmetic average of the Poisson
 and Wigner distributions. There is no degeneracy for both (a) and (b)
 [$P(s)=0$ at $s=0$]. }
\end{figure}

For all the investigated nonintegrable cases except for the case where
$\Delta_2$ is close to $1.0$ and $J_2 \neq 0$, $P(s)$ can be  approximated
by the arithmetic
average of the Poisson and Wigner distributions:
\begin{equation}
 P_{\rm av}(s)=\left[ P_{\rm Poi}(s) + P_{\rm Wig}(s) \right]/2,
\label{eq:average}
\end{equation}
as shown in Fig.~\ref{fig:nn2}. 

\begin{figure}
\includegraphics[width=8cm]{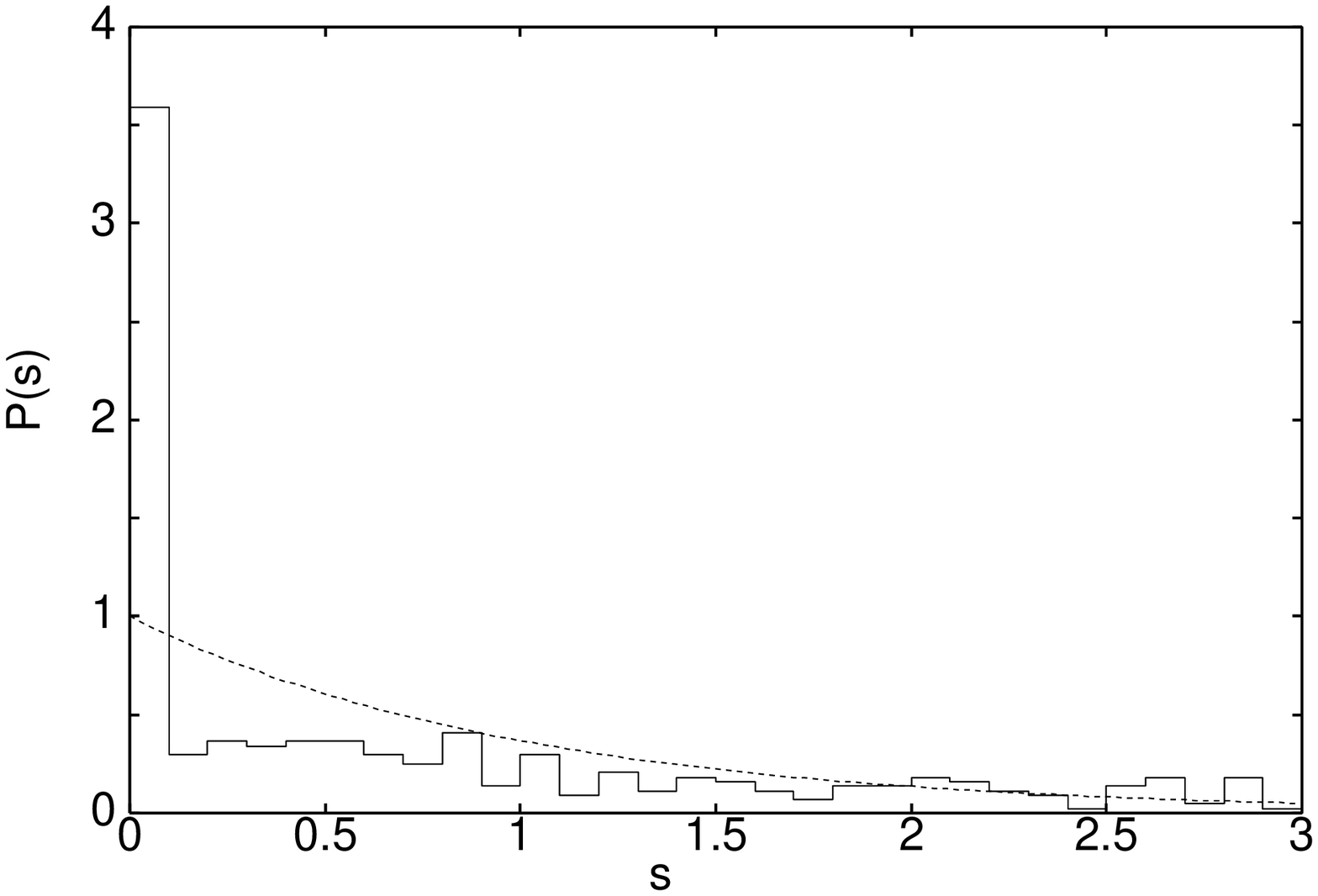}
\caption{\label{fig:in2} Level-spacing distribution of integrable 
 \textit{XXZ} chain ($J_2=0$) for
 $L=16$, $J_1=1$, $\Delta_1 =\cos(\pi/3)=0.5$. Broken line shows the
 Poisson distribution. 
The peak at $s=0$ is given by degenerate eigenvalues.}
\end{figure}

Let us remark on the special case of the integrable \textit{XXZ} chain 
related to a root of unity. 
At the special values of $\Delta_1$, we find that 
 $P(s)$ shows a novel peak at $s=0$ as shown in Fig.~\ref{fig:in2}.  
  For $L=16$,  large degeneracies remain for 
$\Delta_1 = \cos (\pi /i)$, $i=2,3,4$, even after   
the desymmetrization procedure  mentioned in Sec.~\ref{sec:method} is 
performed. 
 We recall that for generic values of $\Delta_1$, 
 no degeneracy remains after the desymmetrization procedure is 
 completed and the Poisson 
distribution is obtained as shown in Fig.~\ref{fig:in}. 
The appearance of the peak in $P(s)$ at $s=0$ should be consistent with 
the $sl_2$ loop algebra symmetry, which holds for the transfer matrices
of the \textit{XXZ} and \textit{XYZ} spin chains only 
at the special  values of $\Delta_1$. \cite{DFM,FM,Deguchi} 
The result could be related to the observation in Ref.~\onlinecite{8vertex} 
that the eigenvalue spacing distribution
$P(s)$ has a peak 
at $s=0$ for the transfer matrix of the eight-vertex model 
under the free-fermion conditions. 
Details should be discussed in forthcoming papers.

\section{\label{sec:discuss}Discussion}

Let us discuss 
some possible reasons why the nointegrable models give 
the non-Wigner distributions. 
We may consider two reasons: extra symmetries  or 
finite-size effects. 

 The non-Wigner distributions, such 
 as shown in Figs.~\ref{fig:nn} and \ref{fig:nn2}, 
suggest that the Hamiltonian should have some extra symmetries. 
It must be a possible standard interpretation 
according to Refs.~\onlinecite{Pals} and \onlinecite{Angles}. 
 However, for the NNN coupled \textit{XXZ} chains, 
it is not clear whether there indeed exist some extra symmetries 
other than reflections, translations, and spin rotations  
 around the $z$ axis.

 We may also consider some finite-size effects, 
 since the distribution $P(s)$ is always obtained
  for some finite systems.  
All the calculations in this paper are performed on 16-site 
chains. Recall that the largest sector of the NNN coupled \textit{XXZ} Hamiltonian 
has 440 eigenvalues. The number is not small. If the level-spacing 
distribution calculated on hundreds-site or thousands-site chains could 
give a Wigner-like distribution, 
then the finite-size effects should be nontrivial.

\begin{acknowledgments}
 The present study was partially supported by the Grant-in-Aid for
 Encouragement of Young Scientists (A): No.~14702012.
\end{acknowledgments}

\end{document}